\documentclass[preprint, tighten, twocolumn]{aastex62}
\usepackage{soul}
\newcommand{\swift}{{\it Swift}}
\newcommand{\hst}{{\it HST}}
\newcommand{\iue}{{\it IUE}}
\newcommand{\et}{et al.\,}
\newcommand{\tvw}{\tau_{V/W2}}

\newcommand{\rae}[1]{{#1}}
\newcommand{\kdh}[1]{{#1}}

\newcommand{\del}[1]{{}}

\shorttitle{IBRM of F9}
\shortauthors{Edelson \et}

\begin{document}

\title{Intensive broadband reverberation mapping of Fairall~9 with 1.8 years of daily Swift monitoring}

\author[0000-0001-8598-1482]{R. Edelson}
\affiliation{Eureka Scientific, Inc., 2452 Delmer Street, Suite 100, Oakland, CA 94602, USA}
\email{email: rickedelson@gmail.com}

\author[0000-0001-6481-5397]{B. M. Peterson}
\affiliation{Retired}

\author[0000-0001-9092-8619]{J. Gelbord}
\affiliation{\rae{Spectral Sciences Inc., 30 Fourth Ave.\ Suite 2, Burlington, MA 01803, USA}}

\author[0000-0003-1728-0304]{Keith Horne}
\affiliation{SUPA Physics and Astronomy, University of St. Andrews, Fife, KY16 9SS, Scotland, UK}

\author{M. Goad}
\affiliation{University of Leicester, Department of Physics and Astronomy, Leicester, LE1 7RH, UK}

\author[0000-0002-0151-2732]{I. McHardy}
\affiliation{School of Physics and Astronomy, University of Southampton, Highfield, Southampton, SO17 1BJ, UK}


\author[0000-0003-4808-092X]{S. Vaughan}
\affiliation{University of Leicester, Department of Physics and Astronomy, Leicester, LE1 7RH, UK}

\author[0000-0001-9191-9837]{M. Vestergaard}
\affiliation{Niels Bohr Institute, University of Copenhagen, Jagtvej 155, 2200 Copenhagen N, Denmark}
\affiliation{Steward Observatory, Dept. of Astronomy, University of Arizona, 933 N. Cherry Ave, Tucson AZ 85721, USA}

\begin{abstract}

We present 1.8 years of near-daily \swift\ monitoring of the bright, strongly variable Type~1 AGN Fairall~9. 
Totaling 575 successful visits, this is the largest such campaign reported to date.
Variations within the UV/optical are well-correlated, with \del{variations at} longer wavelengths lagging \del{those at} shorter wavelengths in the direction predicted by thin disk/lamp-post models. 
The correlations are improved by ``detrending;'' subtracting a second-order polynomial fit to the UV/optical light curves to remove long-term trends that are not of interest to this study.
Extensive testing indicates \del{that} detrending with higher-order polynomials removes too much \del{of the} intrinsic variability signal on reverberation timescales. 
These data provide the clearest detection to date of interband lags within the UV, indicating that neither emission from a large disk nor diffuse continuum emission from the broad-line region can independently explain the full observed lag spectrum.
The observed X-ray flux variations are poorly correlated with those in the UV/optical.
Further, subdivision of the data into four $\sim$160 day light curves shows that the UV/optical lag spectrum is highly stable throughout the four periods, but the X-ray to UV lags are unstable, significantly changing magnitude and even direction from one period to the next.
This indicates \del{an} \rae{the} X-ray to UV relationship \del{that} is more complex than predicted by the simple reprocessing model often adopted for AGN.
A ``bowl'' model (lamp-post irradiation and blackbody reprocessing on a \del{flat} disk with a steep rim) 
\del{fits the observed lags but disagrees with the variable spectral energy distribution to $\sim$11\%.} 
\rae{fit suggests the disk thickens at a distance ($\sim$10 lt-day) and temperature ($\sim8000$K) consistent with the inner edge of the BLR.}
\end{abstract}


\section{Introduction}\label{sec:intro}

It is generally thought that essentially all massive galaxies contain a supermassive black hole (SMBH) in their centers, that is either ``active'' or ``quiescent.''
Active galactic nuclei (AGN) -- which are the most powerful, discrete, quasi-continuous sources of luminosity in the known universe -- are seen in a few percent of bright galaxies, while ``quiescent'' SMBH occur in the vast majority of massive galaxies \citep{Ferrarese00,Gebhardt00} including our own (Sgr A$^*$; \citealt{Genzel96,Ghez98}). 
AGN show a highly luminous, hard ionizing source, with strong excesses in the UV and X-rays, that is often more luminous than the entire underlying galaxy.
AGN also typically exhibit broad emission lines from gas excited by the hard radiation field.
Support for this dichotomy between quiescent and active galactic nuclei comes from the identification of ``changing look'' AGN, cases in which a nuclear continuum and broad emission lines emerge or disappear on short timescales of months or years (e.g., \citealt{Penston84,Denney14,Shappee14,Runnoe16,LaMassa17,MacLeod19,Raimundo19}).
For this reason, it is crucial to understand the physical conditions and processes that give rise to AGN activity driven by SMBH.

A key feature of the active state of SMBH is that AGN are thought to be 
accreting matter onto the SMBH at high rates.
In the current paradigm of AGN, the SMBH is immediately surrounded by a hot, relatively spherically symmetric X-ray-emitting corona, with a larger, flattened disk of optically thick matter accreting inward.
This disk/corona system is thought to be where the bulk of an AGN's large luminosity is produced, ultimately by the liberation of gravitational potential energy as matter heats up as it spirals in towards the SMBH.
The corona and inner disk, which are thought to produce the bulk of the ionizing photons, are thought to have size scales of order 10 and 100 gravitational radii, respectively.
Thus, for a $\sim 10^7 M_\odot $ black hole these sizes would be of order $\sim 0.01$ and 0.1 lt-day \citep{Frank02}.
Outflows from the disk or blobs of accreting matter produce larger regions of ionized gas: the broad-line region (BLR; generally light days to light weeks in size) and narrow-line region (NLR; up to lt-yrs in size).

Unfortunately, with a few exceptions (e.g., 3C\,273 [\citealt{Gravity18}]; M\,87 [\citealt{Event19}];
NGC 1068 [\citealt{Gravity20a}], IRAS 09149$-$6206 [\citealt{Gravity20b}],
NGC 3783 [\citealt{Gravity21}]), the small sizes of AGNs
and their large physical distances from us mean that the angular sizes of the central engines are too small to be imaged directly.
Thus indirect probes, such as source variability, provide some of the strongest constraints on the compact central regions. 
In particular, reverberation mapping has proven to be a particularly powerful technique \citep{Blandford82,Peterson93}. 
Measured lags (or ``light echoes'') of the variable driving continuum in the spatially extended, variable BLR emission line gas have been used to constrain the BLR size, structure and orientation, and to estimate the SMBH mass of over 100 AGN (for recent compilations, see \citealt{Bentz15} and \citealt{DallaBonta20}).
Gravitational lensing (e.g., \citealt{Morgan10}) can also be used to study the structure of the central engines of more distant AGN.

These basic reverberation mapping principles are now also being used to probe the smaller but more physically and energetically important accretion disk/corona region. 
The ``reprocessing'' model predicts strong correlations among the continuum bands, with X-rays leading the UV and the UV leading the optical. 
Although many early campaigns found good interband correlations (e.g., \citealt{Cackett07}), for over a decade, none had unambiguously detected ($> 3 \sigma$) the expected interband lags.
The breakthrough came with the first AGN intensive broadband reverberation mapping (IBRM)\footnote{The previous papers listed in this paragraph used a slightly different acronym, IDRM, for Intensive Disk Reverberation Mapping. However this and other recent campaigns show that the accretion disk is not the only, and in fact may not even be the dominant cause of the observed broadband interband lags. 
Thus we believe this name more aptly describes this technique.} monitoring campaign in 2014, of the target NGC 5548, combining \swift\ \citep{Edelson15}, ground-based optical \citep{Fausnaugh16} and \hst\ \citep{DeRosa15} to blanket the accessible X-ray, UV and optical wavelengths with unprecedented coverage in both temporal and photon frequencies.
Subsequent IBRM campaigns were done on several targets (e.g., NGC~4151 [\citealt{Edelson17}], 
NGC~4593 [\citealt{McHardy18}], 
Mrk~142 [\citealt{Cackett18}], 
Mrk~509 [\citealt{Edelson19}], 
Fairall~9, [F9 hereafter; \citealt{Hernandez20}],
Mrk~110 [\citealt{Vincentelli21}], 
Mrk~817 [\citealt{Kara21}], and
Mrk~335 [\citealt{Kara23}]).

The common feature of all these IBRM campaigns is intensive (daily or sub-daily) monitoring for a total of 200+ visits with \swift\ in the 0.3--10 keV X-rays and six broadband UV/optical filters spanning $\sim 1900 - 5500 $~\AA, simultaneous with ground-based robotic monitoring and often additional observatories (e.g., \hst\ for NGC~4593, \citealt{Cackett20}).
These campaigns yielded three main observational results, which appear to be generally consistent from object to object:
\begin{enumerate}
\item The UV/optical variations are strongly correlated with lags $\tau$ increasing with wavelength $\lambda$ 
consistent with $ \tau \propto \lambda^{4/3}$, as predicted by the standard thin disk model \citep{Shakura73}, although the disk size appear to be larger than predicted by a factor of $\sim$2--3.
\item The $U$ band (which contains the Balmer jump) shows an excess lag relative to this trend, apparently due to diffuse continuum emission (DCE) from the broad-line region.
\item The correlation between X-ray and UV variations is much weaker than the correlations among the UV/optical, casting doubt on the standard ``reprocessing'' model that links the observed X-ray emission (thought to arise in a hot corona immediately surrounding the SMBH) and the UV-emitting accretion disk.
\end{enumerate}

These three observational results challenge various aspects of the current standard model of AGN central engines, and thus have spurred a tremendous amount of theoretical work.
Perhaps the first such theoretical interpretation of IBRM data was \cite{Gardner17}, which utilized a far-UV\rae{(FUV)}-emitting Comptonized reprocessor that lies between the X-ray emitting corona and an accretion disk with a truncated inner edge (at $\sim 70 R_g$, where $R_g=G\,M/c^2$ is the gravitational radius) to explain the observed weak X-ray to UV correlation in the original NGC 5548 campaign.
A similar approach by \cite{Mahmoud20} found that the NGC 4151 IBRM data again required a Comptonized reprocessor, with a highly truncated accretion disk or possibly no disk at all.
\rae{This methodology was also applied by \cite{Hagen23} and \cite{Hagen24} to the initial F9 dataset of \cite{Hernandez20}, finding further evidence of a second FUV component and a truncated disk.}

Alternatively, \cite{Kammoun19,Kammoun21a,Kammoun21b} explained these and other IBRM data with a more conventional geometry, albeit with a hotter disk, \del{larger Eddington ratio,} and \del{a much} larger lamp-post height \del{of $\sim 60 R_g$}.
\cite{Panagiotou22a,Panagiotou22b} expanded on these results by performing the analysis in the temporal frequency domain and simultaneously measuring the power spectral density functions, yielding similar results.

Additionally some theoretical studies focused only on the first two points, based on UV/optical data only.
\cite{Neustadt22} models the UV/optical light curves of seven AGN IRBM campaigns by assuming a standard \cite{Shakura73} disk and a lamp-post illuminator.
They found that the fluctuations consisted of both ingoing waves, along with outgoing waves as would be expected from the standard reprocessing model.
\cite{Netzer22} analyzed UV/optical IBRM light curves of six AGN, concluding that they could be explained by a combination of BLR emission and a standard-sized \cite{Shakura73} disk.
Clearly the IBRM technique, in use for less than a decade, has motivated a wide variety of theoretical models.
This in turn underlines the need for even more constraining IBRM data.

As noted above, \cite{Hernandez20} reported the results of IBRM monitoring of F9 ($ z = 0.047 $, \citealt{Neeleman16}) over a $\sim$9 month period.
\rae{It is worth explicitly noting that a number of studies have already successfully modeled this initial dataset \cite[e.g.][]{Netzer22,Kammoun23,Hagen23,Hagen24}.}
\del{but this target was, in fact,} 
\rae{In fact this target was}
observed by \swift\ on an approximately daily basis during the period MJD 58251.50–58900.50, a total of $\sim$1.8 yr.
This \rae{has} yielded the longest \swift\ AGN IBRM dataset that has been gathered to date, with at least twice the time span and nearly twice \rae{the} total \del{number of} visits of any \del{other} AGN.
\rae{This longer dataset is reported in the current paper.}
Additional monitoring was obtained, i.e., ground-based robotic optical photometry and spectroscopy and {\it NICER} X-ray spectroscopy; those results will be reported elsewhere. 

The intensive F9 monitoring presented herein is by far the largest \swift\ AGN IBRM dataset ever gathered (although a current campaign, on Mrk 817 \citep{Cackett23}, is poised to eventually surpass it).
This allows analyses that were not previously feasible.
With such a large, homogeneous dataset, we can divide the full 1.8 year light curve into subsets that are equivalent in size to individual past IBRM datasets and compare the interband lag spectra at different times and when the target is in different flux states.
An analogous situation exists in emission-line reverberation studies: NGC 5548 has been subject to nearly 20 reverberation campaigns, enabling study of the line response at different times and under a variety of conditions (\rae{see} \citealt{Peterson13,Pei17,DeRosa18}. \del{and refeences therein}\rae{and associated references and citations}).

As the main goal of this paper is to present this new dataset to the community, only a small amount of model testing is done herein.
Instead it is hoped that the aforementioned authors (and others) will utilize these data to test and refine their models, and \rae{perhaps to} develop entirely new ones.
This paper is organized as follows:
Section~2 summarizes the observations and data reduction,  
Section~3 presents the timing analysis, 
Section~4 discusses theoretical implications of these results,
and Section~5 gives some brief concluding remarks.

\section{Observations and Data Reduction}\label{sec:data}

\subsection{Observations}\label{sec:obs}

F9 was the subject of daily\ IBRM monitoring for 1.8 years, covering the period MJD 58251–58901.
Data from two of \swift's three instruments are used in this paper: the X-Ray Telescope \citep[XRT,][]{Burrows05} and the UltraViolet/Optical Telescope \citep[UVOT,][]{Roming05}.
All XRT observations in this campaign were made in photon counting (PC) mode.
All UVOT observations were made using 6-filter modes.  
Initially the workhorse mode 0x30ed was employed, but observations were switched to 0x224c once this mode became available (from MJD 58594 onwards).  
This new UVOT mode is specifically designed for IBRM campaigns, with extra weight given to exposures in the extremal filters $W2$ and $V$.  
In particular the fraction of time on target assigned to $V$ band is increased by a factor of $\sim$3, contributing to clear improvements in the signal/noise ratio of the $V$ light curve (see Section~\ref{sec:uvot}) and $W2$--$V$ lag constraints in the latter half of the campaign. 

Additional multi-wavelength monitoring was obtained with the {\it NICER} X-ray mission and the ground-based robotic Las Cumbres Observatory global telescope network, operating at optical wavelengths.
Furthermore \swift\ collected additional data at lower cadence (once every $\sim$4 days) during the period MJD 58902--59230.
None of these data are analyzed in this paper because here we focus on the analysis of a homogeneous, relatively uninterrupted set of \swift\ light curves.
These other datasets will be the subject of future papers.

\subsection{UVOT data reduction}\label{sec:uvot}

This paper's UVOT data reduction follows the general procedure described by \cite{Hernandez20}.
This process has four steps: definition of the observing wavebands, flux measurement, removal of data points that fail quality checks, and identification and masking of low sensitivity regions of the detector.
Each step is described in turn below.

\begin{figure}[h]
\includegraphics[trim={0.5cm 11.7cm 1cm 2.8cm},clip,width=0.48\textwidth]{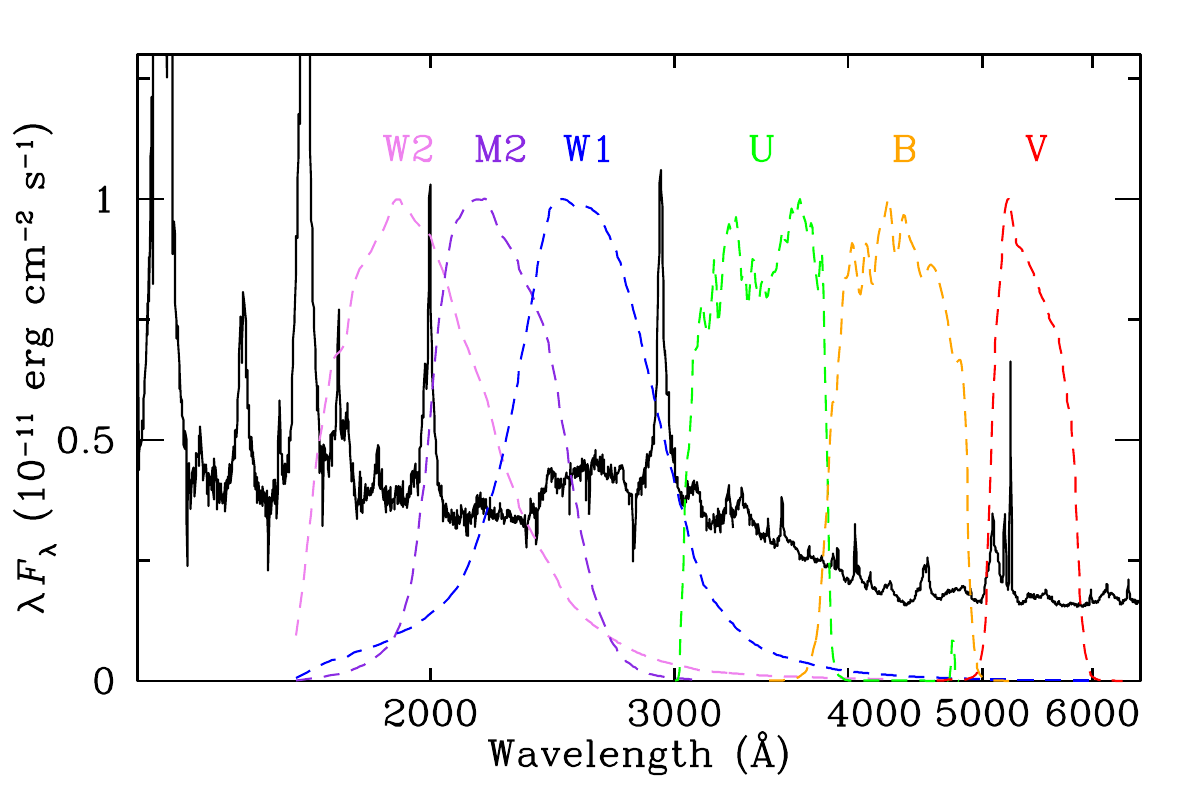} 
\caption{UV/optical spectrum of F9 created by combining \hst\ FOS spectra obtained on 1993 January 21 with a spectrum from the ESO 1.5-m telescope taken on 1994 July 14 \citep{Santos97}.
Spectra are scaled to match in overlapping wavelength regions as the ground-based data were obtained under non-photometric conditions.
The colored, dashed lines show the passband transmission curves of each \swift\ UVOT band.
Note that the names ``W2,'' ``M2,'' and ``W1'' are shorthand for ``UVW2,''
``UVM2,'' and ``UVW1'', as used throughout this paper.}
\label{fig:bands}
\end{figure}

In this work the band centers are defined \citep{Koornneef86} as the ``pivot wavelength,'' 
\begin{equation}
\lambda_p(P) = \sqrt{{\int P(\lambda)\, \lambda \, d \lambda} \over {\int P(\lambda)\, d \lambda / \lambda }} \ ,
\end{equation}
where $P(\lambda)$ is the passband transmission curve. The colored dashed curves in Fig.~\ref{fig:bands} show $P(\lambda)$ for each band.
The associated rms bandwidth is
\begin{equation}
\lambda_{rms} = \sqrt{{\int(\lambda-\lambda_p)^2\, P(\lambda)\, d \lambda} \over {\int P(\lambda) \,d \lambda }} \ .
\end{equation}
This source-independent parameterization has the advantage that it allows an exact conversion between broadband flux densities $f_\nu$ and $f_\lambda$.
Indeed using the pivot point as reference and normalization makes the parameters of any 2-parameter linear regression independent, and thus gives a diagonal Hessian matrix which is trivial to invert.
These quantities are tabulated in Table~\ref{tab:bands} and the UVOT passbands and a spectrum of F9 are shown Fig.~\ref{fig:bands}.

\begin{deluxetable*}{lcccccccc}[h]
\tablecaption{\swift\ Observing bands and basic variability information \label{tab:bands}}
\tablecolumns{8}
\tablehead{
\colhead{(1)} & \colhead{(2)} & \colhead{(3)} & \colhead{(4)} & \colhead{(5)} & \colhead{(6)} & \colhead{(7)} & \colhead{(8)} 
 & \colhead{(9)} \cr
\colhead{Band/} & \colhead{Band} & \colhead{Band} & \colhead{} & \colhead{} & \colhead{Mean Sampling} & \colhead{Mean} & \colhead{} & \colhead{} \cr
\colhead{Filter} & \colhead{Center} & \colhead{Range} & \colhead{Unit} & \colhead{$N$} & \colhead{Rate (days)} & \colhead{Flux} & \colhead{$F_{var}$} & \colhead{Instrument} } 
\startdata
$HX$ & 3.9 & 1.5--10 & kev & 575 & 1.13 & 0.98 & 0.20 & XRT \cr
$SX$ & 0.7 & 0.3--1.5 & kev & 575 & 1.13 & 0.49 & 0.15 & XRT  \cr
$W2$ & 2055 & 1596--2261 & \AA & 479 & 1.35 & 4.89 & 0.25 & UVOT \cr
$M2$ & 2246 & 2018--2474 & \AA & 443 & 1.47 & 4.20 & 0.24 & UVOT \cr
$W1$ & 2580 & 2134--3066 & \AA & 442 & 1.47 & 3.47 & 0.22 & UVOT \cr
$U$ & 3463 & 3192--3739 & \AA & 459 & 1.41 & 2.41 & 0.20 & UVOT \cr
$B$ & 4350 & 4055--4729 & \AA & 512 & 1.27 & 1.40 & 0.17 & UVOT \cr
$V$ & 5425 & 5202--5734 & \AA & 534 & 1.22 & 1.11 & 0.13 & UVOT  
\enddata
\tablecomments{ 
Column 1: Filter/band name used in this paper.
Column 2: Band center.
Column 3: Band range. 
Column 4: The physical unit for the quantity in Columns~2 and 3.
Column 5: Number of good data points in each band ($N$), after removal of dropouts and other censored data.
Column 6: Mean sampling rate for each band, after removal of bad data.
Column 7: Mean flux, in units of $10^{-14}$ erg cm$^{-2}$ s$^{-1}$ \AA$^{-1}$ for UVOT data and ct s$^{-1}$ for XRT data.
Column 8: Fractional variability ($F_{\rm var}$) \cite{Vaughan03}.}
Column 9: \swift\ instrument (XRT or UVOT).
\end{deluxetable*} 

In the next step, we reprocess all data for uniformity (using version 6.28 of {\tt HEASOFT}) and refined their astrometry \citep[following the procedures of][]{Edelson15} before measuring fluxes using {\tt UVOTSOURCE} from the {\tt FTOOLS}\footnote{\url{http://heasarc.gsfc.nasa.gov/ftools/}} package \citep{Blackburn95}.
Details of this instrument are given e.g., by \cite{Poole08}.
The F9 nucleus and numerous selected field stars are measured, the latter in order to allow various checks.
Source photometry is measured in circular extraction regions of 5\arcsec\ radius, while background fluxes were taken from concentric 40\arcsec--90\arcsec\ annuli.
In the $V$ band in particular, the galaxy contributes significantly to the reported source flux (the background annulus lies beyond the host galaxy).
The final flux values are corrected for aperture losses, coincidence losses, large-scale variations in the detector sensitivity across the image plane, and declining sensitivity of the instrument over time.
These corrections include an updated model for the time dependence of UVOT sensitivity, first introduced in Sept.\ 2020, which causes the fluxes reported here to differ systematically from those of \cite{Hernandez20} and earlier papers.  
For details, refer to the calibration document SWIFT-UVOT-CALDB-15-06 (\href{https://heasarc.gsfc.nasa.gov/docs/heasarc/caldb/swift/docs/uvot/uvotcaldb_throughput_06.pdf}{https://\linebreak[0]heasarc.gsfc.nasa.gov/\linebreak[1]docs/\linebreak[1]heasarc/\linebreak[1]caldb/\linebreak[1]swift/\linebreak[1]docs/\linebreak[1]uvot/\linebreak[1]uvotcaldb\_throughput\_06.pdf}).
In general the fluxes reported herein tend to be lower, differing by a few percent from those in \cite{Hernandez20}, with larger disparities of $\sim$5--8\% in the M2 and W1 filters.
These calibration changes should not strongly affect the conclusions of \cite{Hernandez20} because the systematic light curve shifts are negated by the detrending process.

In the third step, the resulting measurements are used for both automated quality checks and to flag individual observations for visual inspection. 
These automated checks include aperture ratio screenings to catch instances of extended point-spread functions (PSFs) or when the astrometric solution is off, and that the exposure time was a minimum of 20~s.
Data are flagged for inspection when the fitted PSFs of either the AGN or several field stars are found to be unusually large or asymmetric, or if fewer than 10 field stars with robust centroid positions are available for astrometric refinement.  
Upon inspection, observations are rejected if there are obvious astrometric errors, doubled or distorted PSFs, or prominent image artifacts (e.g., readout streaks or scattered light) that could affect the AGN measurement.
Note that we adopt a non-standard setting of 7.5\% for the {\tt UVOTSOURCE} parameter {\tt FWHMSIG} because this yields flux uncertainties more consistent with the rms point-to-point scatter in the data \citep{Edelson17}.

\cite{Edelson15} first noted that $\sim10-15$\% of \swift\ UVOT data suffered from ``dropouts,'' anomalously low fluxes, especially in the three UV bands, which occurs when the target falls on low-sensitivity regions in the UVOT detector.
That work, and subsequent IBRM campaigns, used a bootstrap method to identify and filter out such anomalous data.
\cite{Hernandez20} instead utilized the daily \swift\ monitoring of SgrA* to derive a much more comprehensive and stable map of the entire UVOT detector for this purpose, and applied it to identification and filtering of UVOT dropouts.
That improved technique was applied to filter out UVOT dropouts in the current dataset as well.

\subsection{XRT data reduction}\label{sec:xrt}

The XRT data reduction is essentially the same as described by \cite{Edelson17} so it will only be briefly summarized here.
We utilize the standard \swift\ analysis tools\footnote{\url{http://www.swift.ac.uk/user\_objects}} described by \cite{Evans09}\ to produce light curves that are fully corrected for instrumental effects such as pile up, dead regions on the CCD, and vignetting.
The source aperture varies dynamically according to the source brightness and position on the detector. 

We extract the observation times (the midpoint between the start and end times) in MJD for ease of comparison with the UVOT data.
We utilize ``snapshot'' binning, which produces one temporal bin for each continuous spacecraft pointing. 
This is done because these short visits always occur completely within one orbit with one set of corresponding exposures in the UVOT filters.
We generate X-ray light curves covering two bands, the standard soft (0.3--1.5~keV) and hard ( 1.5--10~keV) bands, referred to as $SX$ and $HX$, respectively.
For a detailed discussion of this tool and the default parameter values, see \cite{Evans09}.

\subsection{Light Curves}\label{sec:lcs}

\begin{figure*}
\begin{center}
 \includegraphics[trim=0cm 0cm 0cm 1cm,clip,width=\textwidth] {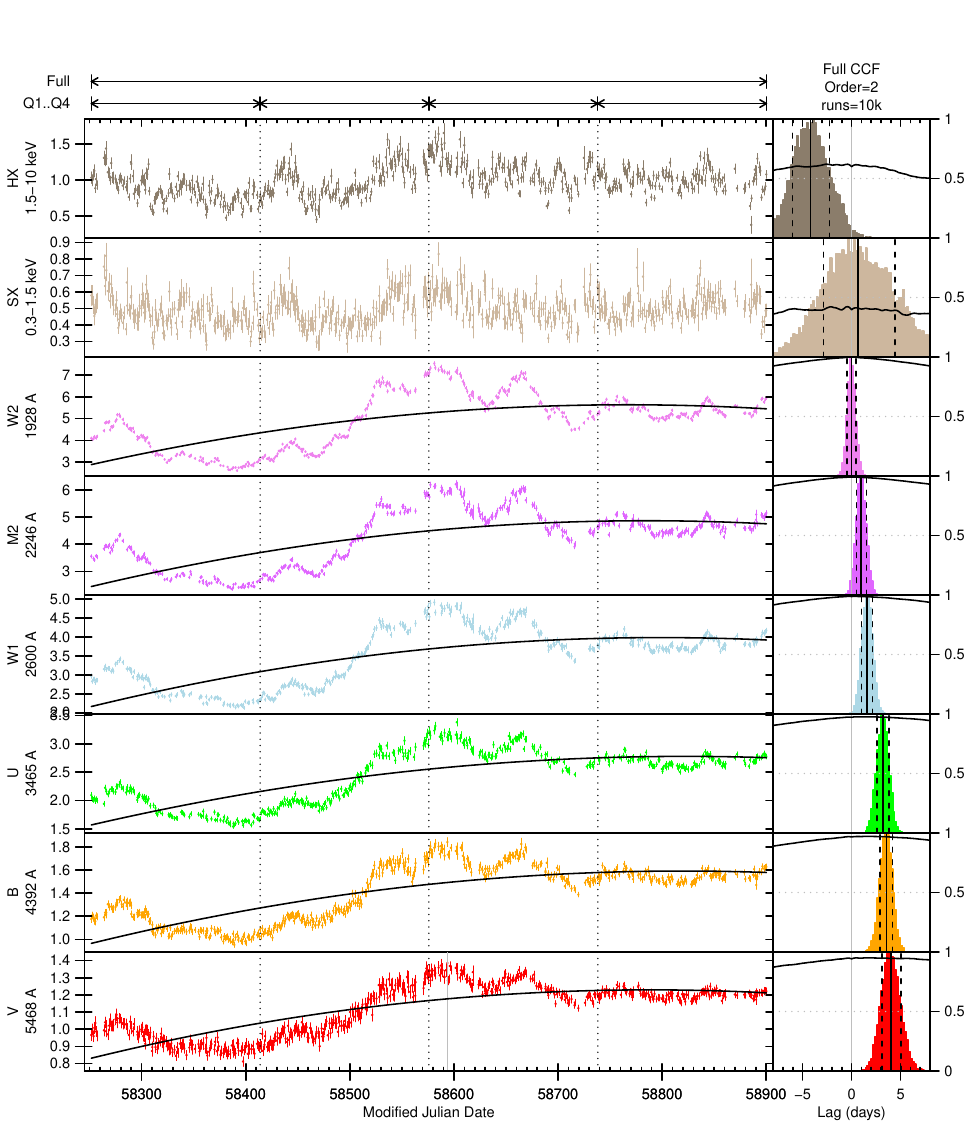}
\caption{{\em Left:} light curves for the intensive \swift\ monitoring campaign.
Units are the same as in Table~\ref{tab:datatab}.
The vertical grey line in the bottom ($V$-band) plot at MJD~58594 denotes the switch to UVOT mode 0x224c, resulting in a considerable improvement in $S/N$ in this band.
The time intervals used in this paper are shown at the top:
the ``Full'' analysis period incorporates all data shown and the four quarterly periods (Q1...Q4) are delimited by dashed lines.
As discussed in Section~\ref{sec:detrend}, the X-ray data were not detrended and UV/optical were detrended by subtracting a second-order polynomial, shown as a black line.
{\em Right:} CCF (black lines) and probability density functions (colored histograms) as a function of time delay (lag) for the full period.
The observing bands are shown in the same colors and order as the light curves. 
Grey vertical lines mark a zero-day lag.
All CCFs are measured relative to the ultraviolet $W2$ band.
The median lag estimates are shown as a solid vertical black line and the 68\% confidence interval as dashed vertical black lines.
\label{fig:lccf}}
\end{center}
\end{figure*}

The resulting light curves are shown in Fig.~\ref{fig:lccf} and tabulated in Table~\ref{tab:datatab}.
The XRT and UVOT observing bands are tabulated in Table~\ref{tab:bands}.

\begin{center} \begin{deluxetable}{lccc}
\tablecaption{\swift\ F9 IBRM Campaign Data\label{tab:datatab}}
\tablewidth{4in} \tablecolumns{4}
\tablehead{
\colhead{\hspace{1cm}(1)\hspace{1cm}} & \colhead{\hspace{1cm}(2)\hspace{1cm}} & \colhead{\hspace{1cm}(3)\hspace{1cm}} & \colhead{\hspace{1cm}(4)\hspace{1cm}}  \cr
\colhead{Filter} & \colhead{MJD} & \colhead{Flux} & \colhead{Error} } 
\startdata
$W1$ & 58251.6563 & 2.970 & 0.062 \cr
$U$  & 58251.6570 & 2.096 & 0.052 \cr
$B$  & 58251.6574 & 1.175 & 0.031
\enddata
\tablecomments{Column 1: Filter/band used to measure the broadband flux and error.
Column 2: Modified Julian Day at the midpoint of the exposure.
Column 3: Mean flux of the data point.  UVOT fluxes are given in units of  $10^{-14}$ erg cm$^{-2}$ s$^{-1}$ \AA$^{-1}$ and X-ray fluxes  in units of ct s$^{-1}$.
Column 4: Uncertainty on the flux, in the same units as Column~3.
Only a portion of this table is shown here to demonstrate its form and content. 
A machine-readable version of the full table is available online.}
\end{deluxetable} \end{center}

\section{Analysis}\label{sec:tsa}

\subsection{CCF Methodology}\label{sec:ccf1}

Cross-correlation functions (CCFs) were constructed using the interpolated cross-correlation function method \citep{Gaskell87}.
We used the {\tt sour} code\footnote{This code is available at \url{https://github.com/svdataman/sour}}, which is based on the specific implementation presented by \cite{Peterson98} and modified by \cite{Peterson04}.
In computing the cross-correlation functions, the mean and standard deviations for each lag value were derived ``locally,'' i.e., using only the portions of the light curves that overlap in time for a given test lag.
Because this time window is fixed the number of overlapping points will change by a small amount between different lags.
As noted in point 2 below, we limit the half-widths of the lag searches to $\sim$1/3 of the total light curve duration, meaning that at least $\sim$1/3 of the data points do not participate in the measurement of $r$ at any $\tau$.
As discussed by \cite{White94}, local rather than global normalization is appropriate for non-stationary AGN light curves.

Due to the uneven sampling of the light curves, we use interpolation to perform the CCF analysis. 
The CCF is computed in both directions, once with the reference light curve leading and once with the subsidiary light curve leading. 
The $W2$ light curve is always the reference and the other bands are considered to be the subsidiary bands in this analysis.
W2 band is chosen because it has the shortest UV wavelength and thus is closest to the thermal peak of the accretion disk, and because it shows the strongest fractional variability of all the UV/optical bands (see Table~\ref{tab:bands}).

We then use the ``flux randomization/random subset selection'' (FR/RSS) method \citep{Peterson98,Peterson04} to estimate uncertainties on the measured lags.
This is a Monte Carlo technique in which lags are measured from multiple realisations of the CCF.
The FR aspect of this technique perturbs in a given realization each flux point by a random Gaussian deviate whose width is based on the quoted uncertainty associated with that data point.
In addition, for a time series with $N$ data points, the RSS randomly draws with replacement $N$ points from the time series to create a new time series.
In that new time series, the data points selected more than once have their error bars decreased by a factor of $n_\mathrm{rep}^{-1/2}$, where  $n_\mathrm{rep}$ is the number of repeated points. 
Typically a fraction of $(1-[1/N])^N\rightarrow 1/e$ of data points are not selected for each RSS realization. 
In this paper, the FR/RSS is applied to both the reference and subsidiary light curves in each CCF pair.
The CCF is then measured and a lag determined to be the weighted mean of all points with $ r > 0.8\, r_\mathrm{max} $, where $ r_\mathrm{max} $ is the maximum value obtained for the correlation coefficient $r$ \citep[e.g.,][]{Edelson19}. 
Implementation of the FR/RSS version of the interpolation-based cross-correlation method requires specifying a number of parameters:

\begin{enumerate}
    \item The interpolation step size. 
    This should be smaller than the typical spacing between data points in the light curves. 
    However, using too fine a step size results in artifacts on the shortest timescales because points midway between real data points are highly correlated with the adjacent interpolated points, thus producing artificially high local maxima in the CCF on timescales of half the step size. 
    We find that a step size of 0.2 days works well for these data.
    \item The range of lags to be explored. 
    As the light curves are shifted away from one another, an increasing number of points at the ends of a light curve has no counterpart in the other time-shifted spectrum. 
    Thus the number of points that contribute to the CCF computation decreases with increasing lag. 
    To avoid edge effects resulting from a smaller data sample, we set the lag range to a value around 1/3 the length of the light curve being analyzed.
    This yields a lag range of 216 days for the ``full'' sample analyzed in Section~\ref{sec:full}, and 54 days for the  ``quarterly'' dataset analyzed in Section~\ref{sec:quarterly}.
    \item The number of Monte Carlo realizations should be large enough that the resulting probability density functions, or cross-correlation centroid distributions (CCCDs; e.g., the shaded regions on the right side of Fig.~\ref{fig:lccf}) should be smooth, with all features due to the data/method themselves, and not to small-number statistics.
    For these data, we find that 10,000 realizations met this criterion.
\end{enumerate}


\subsection{CCF Analysis of the Full Dataset}\label{sec:full}

We first consider a cross-correlation analysis of light curves from the full-period data, from MJD~58251 through MJD~58900. 
The CCFs and CCCDs, measured relative to the \swift\ $W2$ band, are shown for all the bands in the right-hand column of Fig.~\ref{fig:lccf}, with the corresponding light curves in the left-hand panel.
The CCFs are shown as solid curves and the CCCDs, from the FR/RSS analysis are shown as colored histograms.
Correlated variations are evident in the light curves, and the response of each band is highly localized to a several-day window near zero.
Table~\ref{tab:full} presents the measured interband lags and 68\% confidence limits from the CCCDs.

The UV/optical bands are all highly correlated,
with $r_{max} \ge 0.95$, and
the UV/optical lags are well-enough measured to reveal a clear increase with wavelength. 
In contrast, the X-ray bands are only weakly correlated with $W2$, with $r_{max} < 0.7$, and
the X-ray lags have much larger uncertainties.
While the general shape of the UV/optical variations is evident in the X-ray light curves, the faster X-ray variations are absent or much weaker in the UV/optical light curves. The flare near MJD~58440 is stronger in the X-rays than in the UV/optical.
Similar differences in the X-ray vs UV/optical behavior have been seen in all previous AGN IBRM campaigns, suggesting that the physical process(es) dominating the UV/optical variations are not simply related to (or driven by) the X-ray variability.

\begin{center}
\begin{deluxetable*}{lccccccccccc}
\tablecaption{Full Period Interband Lags for the Zero\del{i}th and Second Orders \label{tab:full}}
\tablewidth{\textwidth}
\tablehead{
& \multicolumn{4}{c}{No detrending (Order = 0)} &
& \multicolumn{4}{c}{Parabolic detrending (Order = 2)} \\
\cline{2-5} \cline{7-10}
\colhead{Band} & 
\colhead{$r_{\rm max}$} & \colhead{$\tau$}& \colhead{$\langle \sigma \rangle$} & \colhead{$S$} && 
\colhead{$r_{\rm max}$} & \colhead{$\tau$}& \colhead{$\langle \sigma \rangle$} & \colhead{$S$} 
\\
\colhead{(1)} & \colhead{(2)} & \colhead{(3)} & \colhead{(4)} & \colhead{(5)} 
&& \colhead{(6)} & \colhead{(7)} & \colhead{(8)} & \colhead{(9)}
}
\startdata
$HX$  &  0.67 & $1.80^{+1.91}_{-2.07}$ & $1.99$ & $0.90$ && 0.62 & $-4.20^{+1.96}_{-1.84}$ & $1.90$ & $-2.21$ \cr
$SX$  &  0.42 & $12.74^{+5.37}_{-3.84}$ & $4.61$ & $2.76$ && 0.42 & $0.63^{+3.79}_{-3.51}$ & $3.65$ & $0.17$ \cr
$W2$  &  1.00 & $0.01^{+0.75}_{-0.78}$ & $0.76$ & $0.01$ && 1.00 & $0.00^{+0.46}_{-0.45}$ & $0.46$ & $-0.01$ \cr
$M2$  &  0.99 & $1.60^{+0.86}_{-0.92}$ & $0.89$ & $1.81$ && 0.99 & $0.98^{+0.54}_{-0.50}$ & $0.52$ & $1.89$ \cr
$W1$  &  0.99 & $1.87^{+1.06}_{-0.97}$ & $1.01$ & $1.84$ && 0.99 & $1.57^{+0.55}_{-0.56}$ & $0.55$ & $2.83$ \cr
$U$  &  0.97 & $4.82^{+1.48}_{-1.48}$ & $1.48$ & $3.26$ && 0.98 & $3.18^{+0.62}_{-0.61}$ & $0.61$ & $5.20$ \cr
$B$ &  0.97 & $3.10^{+1.54}_{-1.52}$ & $1.53$ & $2.02$ && 0.97 & $3.55^{+0.63}_{-0.65}$ & $0.64$ & $5.52$ \cr
$V$  &  0.95 & $1.78^{+2.28}_{-2.43}$ & $2.35$ & $0.76$ && 0.95 & $4.01^{+1.01}_{-0.93}$ & $0.97$ & $4.14$
\enddata
\tablecomments{
Column~1: band/filter name.
Column~2: the maximum value ($r_{\rm max}$) of the cross-correlation coefficient $r(\tau)$. 
Column~3: the median lag ($\tau$) with the asymmetric 68\% confidence interval ($\sigma_{\rm upper}$ and $\sigma_{\rm lower}$) from the CCCDs shown in Fig.~\ref{fig:lccf}.
Column~4: average uncertainty on the median lag, $\langle \sigma \rangle =
\left[ \left(\sigma_{\rm upper}^2 + \sigma_{\rm lower}^2\right)/2 \right]^{1/2}$.
Column~5: the significance, $S \equiv \tau/\langle\sigma\rangle$.
Columns~6--9 are the same as Columns~2--5, but for the Order 2 detrended light curve.
The first two rows give results for hard and soft X-ray (XRT) and the last six for UV/optical (UVOT) filters. The third row ($W2$) is for the autocorrelation instead of cross-correlation.
}
\end{deluxetable*} \end{center}

\subsection{Detrending the UV/optical light curves}\label{sec:detrend}

Examination of Fig.~\ref{fig:lccf} suggests that while the UV/optical light curves are dominated by variations on long timescales, the X-rays are dominated by more rapid variations. 
To quantify this, we construct a simplified power spectral density (PSD) function by first resampling these unevenly sampled data to a regular grid, computing the periodogram by discrete Fourier transform (DFT), and then fitting a simple power law ($P(f)\propto f^{\alpha}$) plus noise model.
This confirms that the UV/optical power spectra are quite steep, with power-law indices in the range $-2.0 \leq \alpha \leq -2.6$. 
Similar behavior is seen in other AGN, especially in observations made with Kepler, which has orders of magnitude faster time sampling. 
In the UV/optical, AGN have steep ``red'' power spectra with values of the spectral index  in the range $-2$ to $-4$ (e.g., \citealt{Edelson14, Vaughan16, Aranzana18}).
The X-ray power spectra, however, are much flatter, with
$-1.0 \leq \alpha \leq -1.2$.
Note that a more detailed analysis that operates in the time domain would avoid the resampling and DFT steps; that will follow in a future contribution.

In the case of the UV/optical light curves, the fact that so much of the variability power is on timescales much longer than the reverberation timescales we are investigating here could compromise accurate lag measurement. 
In order to reduce the risk of bias, many previous investigations sought to reduce the low-frequency power by ``detrending'' the light curves \citep{Welsh99}, i.e., subtracting relatively slow variability signals in order to isolate or enhance the rapid variability that is most suited to measuring short lags (e.g., \citealt{Denney10,Li13,Peterson14,McHardy14,Li18,McHardy18,Zhang19,Raiteri21,Rakshit20,Hernandez20,Li22,Beard23}).

There is, however, no standard prescription for how to detrend light curves; detrending is generally performed on an {\em ad hoc} basis, which may sometimes lead to controversial, or in the worst case, contradictory results.
As an example, consider the case of NGC 7469 which was observed with the {\it Rossi X-Ray Timing Explorer (RXTE)} in the 2--10\,keV range and with the {\it International Ultraviolet Explorer (IUE)} in the 1150--1975\,\AA\ range in 1996. 
This was the most intensive AGN monitoring program that predates \swift, with nearly continuous monitoring for a month. 
Initial CCF analysis with no detrending found that the X-rays {\it lagged} the 1315~\AA\ continuum by $\sim$4 days \citep{Nandra98,Nandra00}, while reanalysis by \cite{Pahari20} showed that the heavily detrended X-rays  {\it led} the UV by $\sim$0.3 days.
It is not desirable that contradictory results be obtained from the same dataset based solely on details of how a user prepares the data for analysis.
Thus we have constructed a methodology to perform an ``optimal'' detrending  without allowing the user to set the input parameters (and thus to determine the resultant interband lags).

The high quality of the F9 light curves allows a systematic investigation of how to optimally set detrending parameters, although once the methodology is constructed it can be applied to any IBRM dataset. 
We first explore detrending the UV/optical data by subtracting a polynomial of the form 
\begin{equation}
\label{eq:poly}
    F(t) = \sum_{i=0}^{N} a_i\, (t-t_0)^i,
\end{equation}
where $N$ is the order of the polynomial, $t_0$ is the approximate midpoint of the observations (MJD 58576) and $F(t)$ is treated as the ``background'' flux at time $t$, and $a_i$ are the best-fit coefficients.
Equation~(\ref{eq:poly}) was fitted to each of the UV/optical light curves by linear regression, minimizing the unweighted sum of the squared residuals.
The resulting fits are subtracted from the original light curves and the cross-correlation analysis is performed.
A wide range of polynomial orders is used: 0--8, 10, 12, 14, and 16.
(Order 0 is the case where no detrending is performed.)
Note that detrending is performed only on the steep PSD UV/optical light curves.

\begin{figure*}
\begin{center}
 \includegraphics[trim=0cm 0cm 0cm 0cm,clip,width=\textwidth] {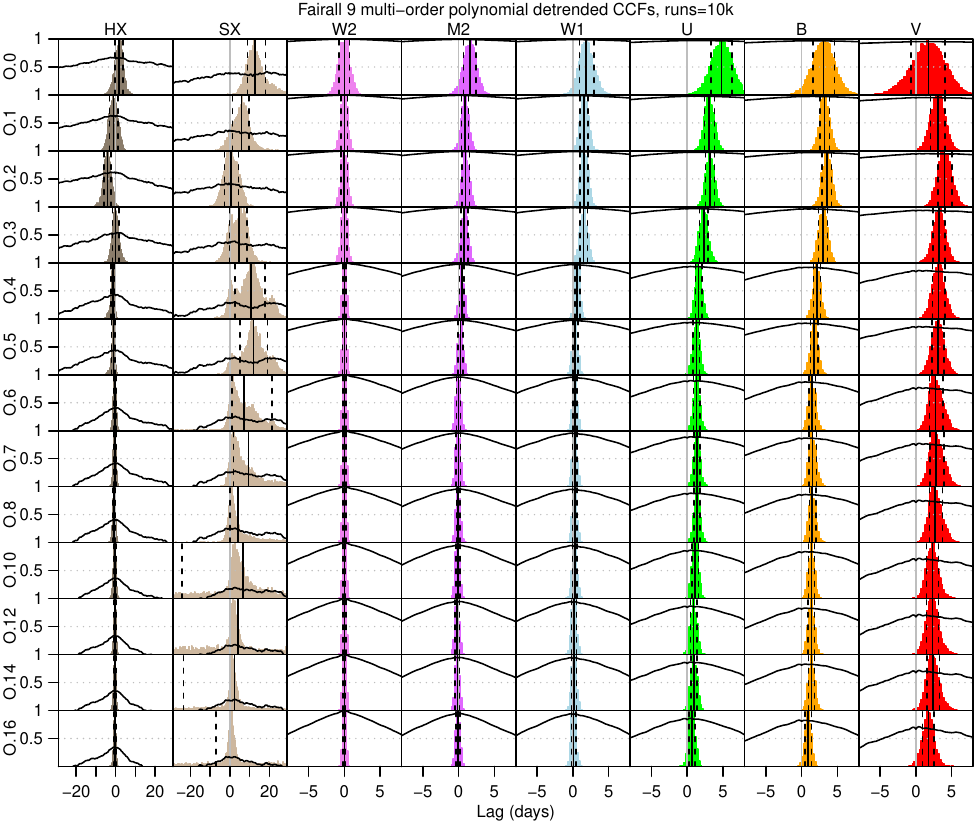}
\caption{Cross-correlation functions (CCFs, black curves) and cross-correlation centroid distributions
(CCCDs, colored histograms) as a function of lag for the full period,
as in the right-hand column of Fig.~\ref{fig:lccf}.
All CCFs are measured relative to the \swift\ $W2$ band.
The median lag estimates are shown as a solid vertical black line and the 68\% confidence interval as dashed vertical black lines.
Each row shows the CCFs and CCCDs for all bands, with the
UV/optical light curves detrended by a polynomial of
order $N$ (Equation~\ref{eq:poly}) indicated in the left-hand legend.
Note the different x-axis scales: $\pm$30 days for the X-ray bands on the left and $\pm$7 days for the UV/optical bands on the right.
\label{fig:detrend}}
\end{center}
\end{figure*}

The resulting CCFs, FR/RSS histograms and 
68\% confidence intervals are shown in Fig.~\ref{fig:detrend}.
We define a figure of merit to be the ``significance'' of the lag measurement: $ S \equiv \tau/\sigma $, e.g., the ratio of Columns (3) to (4), (6) to (7), and (11) to (12) in Table~\ref{tab:full}.
The significance $S$ is tabulated in 
Table~\ref{tab:detrend} for all bands and for light curves detrended with polynomials of order 0 through 16.
The last column of this table shows the average of the five previous columns, that is, the UV/optical CCFs, excluding the $W2$ ACF (which will always be close to zero) and the X-ray CCFs, which show much weaker correlations.
Note that the average value of the figure of merit $S$ is highest for Orders~1--3, with the highest value for Order~2.

The results summarised in Table~\ref{tab:detrend} indicate that some detrending does clearly improve the significance of the lag detections, as the Order~0 (no detrending) values are quite low.
However, note also that the significance then declines for Order~3 and higher.
Examination of Fig.~\ref{fig:detrend} shows why: the lag uncertainties ($\sigma$) become marginally smaller at higher orders, but the lags ($\tau$) get smaller even faster.
This is because overly aggressive detrending removes not only an unrelated confusing signal but also part of the actual lag signal we are trying to measure.
This underscores the importance of being very conservative about performing detrending with anything higher order than, say, a second-order (parabolic) function, or similarly, using a smoothing function with a kernal significantly smaller than the total duration of the experiment.
Based on these results, we utilize a detrending polynomial of order 2 for the UVOT data throughout the rest of this paper.

Note that \cite{Hernandez20} studied the first $\sim$10 months of these dataset, including ground-based LCO {\it uBgVriz} data, through MJD 58550 (see Fig.\ 1 of that paper).
Those data are dominated by a single low temporal frequency parabolic feature covering the full duration studied in that paper.
They detrended this feature and measured interband correlations within it, finding the opposite lag dependence than in the rapid variations (longer wavelength leading shorter wavelengths).
Unsurprisingly, this parabola did not persist into the current analysis, as it has often been noted that the low temporal frequency features seen in red-noise AGN light curves should be viewed with skepticism (see e.g. \citealt{Press78} and subsequent works).

\begin{center}
\begin{deluxetable*}{lccccccccc}
\tablecaption{FR/RSS significance for each band/polynomial order
\label{tab:detrend}}
\tablehead{ 
\colhead{(1)} & \colhead{(2)} & \colhead{(3)} & \colhead{(4)} & \colhead{(5)} & \colhead{(6)}  & \colhead{(7)}  & \colhead{(8)}  & \colhead{(9)}  & \colhead{(10)} \\
\colhead{Order} & \colhead{$HX$} & \colhead{$SX$} & \colhead{$W2$} & \colhead{$M2$} & \colhead{$W1$} & \colhead{$U$} & \colhead{$B$} & \colhead{$V$} & \colhead{Mean} }
\startdata
0 & 0.9 & 2.76 & 0.01 & 1.81 & 1.84 & 3.26 & 2.02 & 0.76 & 1.62 \cr
1 & $-0.46$ & 1.44 & 0.00 & 1.75 & 2.78 & 4.51 & 4.95 & 3.07 & 2.84 \cr
2 & $-2.21$ & 0.17 & $-0.01$ & 1.89 & 2.83 & 5.2 & 5.52 & 4.14 & 3.26 \cr
3 & $-0.01$ & 1.06 & 0.00 & 1.58 & 2.85 & 3.83 & 4.84 & 3.92 & 2.84 \cr
4 & $-1.53$ & 1.42 & 0.00 & 1.35 & 1.8 & 3.52 & 4.34 & 3.64 & 2.44 \cr
5 & $-1.41$ & 1.74 & 0.00 & 0.82 & 0.87 & 2.78 & 3.56 & 3.57 & 1.93 \cr
6 & $-0.66$ & 0.71 & 0.00 & $-0.03$ & 0.99 & 3.01 & 3.08 & 2.89 & 1.66 \cr
7 & $-0.72$ & 0.26 & 0.00 & $-0.12$ & 0.86 & 3.14 & 3.06 & 2.91 & 1.64 \cr
8 & $-1.14$ & 0.14 & 0.01 & $-0.3$ & 0.85 & 3.31 & 3.3 & 2.95 & 1.68 \cr
10 & $-0.66$ & 0.11 & 0.00 & $-0.4$ & 0.82 & 2.55 & 3.43 & 2.82 & 1.54 \cr
12 & $-0.49$ & 0.04 & 0.00 & $-0.64$ & 0.52 & 1.98 & 3.14 & 2.7 & 1.28 \cr
14 & $-0.72$ & 0.04 & 0.00 & $-0.6$ & 0.54 & 2.08 & 3.26 & 2.68 & 1.33 \cr
16 & $-0.73$ & 0.35 & 0.00 & $-0.31$ & 0.32 & 1.5 & 2.1 & 2.19 & 0.97
\enddata
\tablecomments{Column 1: order of the polynomal fit used for detrending. Order zero means no detrending was performed
Columns~2--9: The figure of merit ($\tau/\sigma)$ for each CCF between the listed band and $W2$.
Column~4 refers to the $W2$ band autocorrelation function.
Column~10: average of Columns 5--9, that is, all of the UVOT data except the $W2/W2$ ACF.
}
\end{deluxetable*}
\end{center}

\subsection{Quarterly Analysis}\label{sec:quarterly}

\begin{figure*}
\begin{center}
\includegraphics[width=\textwidth]{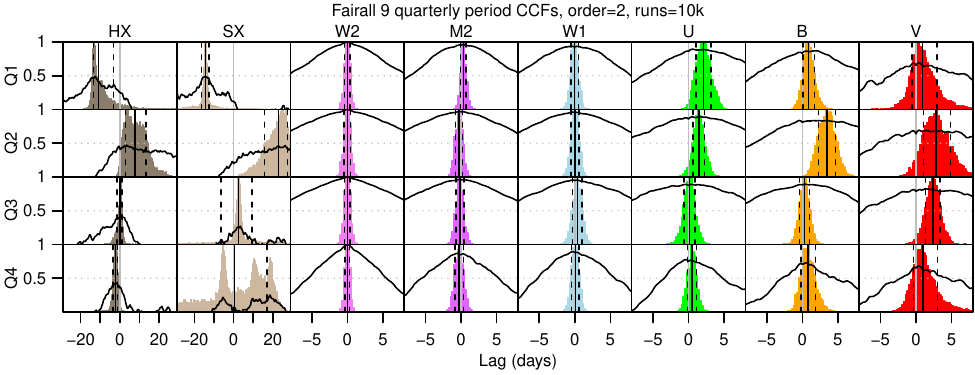} 
\caption{
Plot of CCF (black lines) and probability density functions (colored histograms, corresponding to the colors of the light curves in Fig.~\ref{fig:lccf}) as a function of lag for the four quarterly periods.
The first quarter is on top and the last on the bottom.
Each row shows the CCFs for a different quarter, measured relative to the ultraviolet $W2$ band.
The median lag estimates are shown as a solid vertical black line and the 68\% confidence interval as dotted vertical black lines.
Note the different x-axis scales: $\pm$30 days for the HX/SX bands and $\pm$7 days for the six UVOT bands.
\label{fig:quarterly}}
\end{center}
\end{figure*}

In this section, we perform a novel analysis to compare the stability of lags measured between the UV and X-rays with those measured within the UV/optical.
The $\sim$650 day campaign is subdivided into four equal $\sim$162 day periods, denoted as Q1 through Q4, as listed in Table~\ref{tab:quarterly}.
Each of these show sufficient variability to measure interband correlations and lags (see Fig.~\ref{fig:lccf}).  For the light curves in each quarter,
interband lags are measured as described in Section \ref{sec:ccf1}.
The resulting CCFs are shown in Fig.~\ref{fig:quarterly} and listed in Table~\ref{tab:quarterly}.
The results are stacked by quarter to facilitate comparison.
The right-hand part of Fig.~\ref{fig:quarterly} shows that the lags within the UV/optical are all well-behaved: for each band the medians measured in all four quarters agree to within 1$\sigma$, typically with $r_{max}>0.9$.
This indicates that it is safe to detrend the light curves when measuring lags within the UV/optical.

The left-hand side of Fig.~\ref{fig:quarterly} shows that the lags between the reference $W2$ band and the X-rays are not consistent; from quarter to quarter they do not agree and they are also generally inconsistent with the UV/optical lags.
It has often been noted \cite[e.g.][]{McHardy18,Edelson19} that the UV/optical lag spectra of IBRM AGN generally follow the same ($\tau \propto \lambda^{4/3}$) relation but that the UV/X-rays correlations and lags are inconsistent, but this is the first time this relation has been documented in the same object.
This result also supports the decision in the previous section not to detrend the X-ray light curves.

\begin{center} 
\begin{deluxetable*}{l|ccc|ccc|ccc|ccc}
\tablecaption{Interband Lags for Quarterly Periods.
\label{tab:quarterly}} \tablewidth{0pt} \tablecolumns{13}
\tablehead{
 \colhead{(1)} & \colhead{(2)} & \colhead{(3)} & \colhead{(4)} & \colhead{(5)} & \colhead{(6)} & \colhead{(7)}  & \colhead{(8)}  & \colhead{(9)} & \colhead{(10)} & \colhead{(11)} & \colhead{(12)} & \colhead{(13)} \cr
 \colhead{Band/} & \multicolumn{3}{|c}{Q1 Lags (days) } & \multicolumn{3}{|c}{Q2 Lags (days) } & \multicolumn{3}{|c}{Q3 Lags (days)} & \multicolumn{3}{|c}{Q4 Lags (days) } \cr
\colhead{Filter} & 
\multicolumn{1}{|c}{$r_{\rm max}$} & \colhead{Median} & \colhead{Conf. Int.} & 
\multicolumn{1}{|c}{$r_{\rm max}$} & \colhead{Median} & \colhead{Conf. Int.} & 
\multicolumn{1}{|c}{$r_{\rm max}$} & \colhead{Median} & \colhead{Conf. Int.} & 
\multicolumn{1}{|c}{$r_{\rm max}$} & \colhead{Median} & \colhead{Conf. Int.} 
} 
\startdata
$HX$  &  0.49 & $-10.8$ & $(-13.5,-3.3)$ & 0.47 & $7.8$ & $(2.9,13.4)$ & 0.42 & $0.1$ & $(-1.5,0.9)$ & 0.43 & $-2.2$ & $(-3.6,-1.2)$ \cr
$SX$  &  0.49 & $-14.3$ & $(-16.3,-12.5)$ & 0.46 & $22.8$ & $(15.6,27.6)$ & 0.27 & $2.6$ & $(-6.5,9.3)$ & 0.25 & $0.4$ & $(-29.3,17.0)$ \cr
\hline
$W2$  &  1.00 & $0.0$ & $(-0.4,0.4)$ & 1.00 & $0.0$ & $(-0.5,0.5)$ & 1.00 & $0.0$ & $(-0.4,0.4)$ & 1.00 & $0.0$ & $(-0.4,0.3)$ \cr
$M2$  &  0.96 & $0.3$ & $(-0.1,0.7)$ & 0.97 & $-0.3$ & $(-0.8,0.2)$ & 0.96 & $-0.2$ & $(-0.7,0.4)$ & 0.87 & $-0.3$ & $(-0.9,0.4)$ \cr
$W1$  &  0.94 & $0.0$ & $(-0.5,0.5)$ & 0.96 & $0.0$ & $(-0.5,0.6)$ & 0.96 & $0.4$ & $(-0.3,1.0)$ & 0.89 & $0.1$ & $(-0.4,0.6)$ \cr
$U$  &  0.89 & $2.1$ & $(1.0,3.1)$ & 0.89 & $1.4$ & $(0.7,2.3)$ & 0.89 & $0.1$ & $(-0.7,0.9)$ & 0.78 & $0.5$ & $(-0.1,1.1)$ \cr
$B$  &  0.87 & $0.9$ & $(0.1,1.7)$ & 0.84 & $3.5$ & $(2.3,4.7)$ & 0.89 & $0.3$ & $(-0.4,1.0)$ & 0.73 & $0.8$ & $(-0.1,1.9)$ \cr
$V$  &  0.69 & $0.9$ & $(-0.5,2.9)$ & 0.69 & $2.8$ & $(1.0,4.8)$ & 0.83 & $2.4$ & $(1.3,3.4)$ & 0.68 & $1.0$ & $(-0.4,3.0)$
\enddata
\tablecomments{
Measured interband lag analysis results measured in the four time periods, with all CCFs performed relative to the hardest \swift\ UV band ($W2$).
Column 1: band/filter name.
Columns 2-4: for the quarter 1 (Q1) analysis, these give the maximum value of $r$, followed by the median lag and 68\% confidence interval measured with the FR/RSS technique.
Columns 5-7, 8-10, and 11-13 give the same information as Columns 2-4, but for Q2..Q4, respectively.
The first two rows give the hard and soft X-ray (XRT) results and the last six give those for the six UV/optical (UVOT) filters, so the third row ($W2$) is for the autocorrelation instead of cross-correlation.}
\end{deluxetable*} 
\end{center}

Another interesting feature of this analysis is the relation between the duration of light curves examined and the derived interband lags.
The largest UV/optical wavelength interval, and thus the largest expected interband lag, is between the UVOT bands W2 and V.
For the full 650-day campaign, $\tvw=4.0$~days (Table~\ref{tab:full}), and the mean of the quarterly 162.5-day segments is $\tvw=1.8$~days (Table~\ref{tab:quarterly}).
For the initial 300-day campaign, \cite{Hernandez20} finds $\tvw=2.2$~days.
That is, $\tvw$ increases with duration, albeit with substantial errors and dispersion.

Such behavior is not a surprise given that AGN light curves show ``red-noise’’ variability over a wide range of timescales dominated by the longest timescales sampled, combined with the fact that the expected transfer functions are not delta functions but rather quite broad and asymmetric.
This means that duration light curves will sample preferentially longer lags in the transfer function, or more distant parts of the transfer function.
Thus the ``lag’’ is not merely a function of the physical reprocessor, but also of the driving light curve and the sampling pattern, and we urge caution in interpreting lags as completely well-determined values, independent of the details of the campaigns in which they are measured.

\subsection{Flux-Flux Analysis} \label{sec:fluxflux}

The aim of a flux-flux analysis is to use the observed variations to decompose the AGN light into constant (host galaxy) and variable (accretion disk and/or BLR) components.
This method was developed following the \cite{Winkler92} finding that 
linear relationships hold among observed fluxes in different photometric bands as the AGN brightens and fades. Thus AGN variations are well described by one spectral energy distribution (SED) for the variable component, and another for the constant component, which includes, for example, starlight from the host galaxy \citep{Winkler92}.
We model the observed fluxes $F(\lambda,t)$ versus a dimensionless light curve $X(t)$ that describes the shape of the AGN brightness variations. 
This method has been used in the analysis of data from several previous IBRM campaigns, including the AGN STORM campaign on NGC~5548 \citep{Starkey17}, and the first year of F9 monitoring \citep{Hernandez20}.
The results demonstrate the adequacy of the linear model to describe the UV and optical variations, by up to factors of 5 in the brightness in the case of NGC~5548.

\begin{figure*}[h]
\begin{center}
 \includegraphics[trim=15mm 5mm 55mm 20mm,clip,width=0.49\textwidth]
 {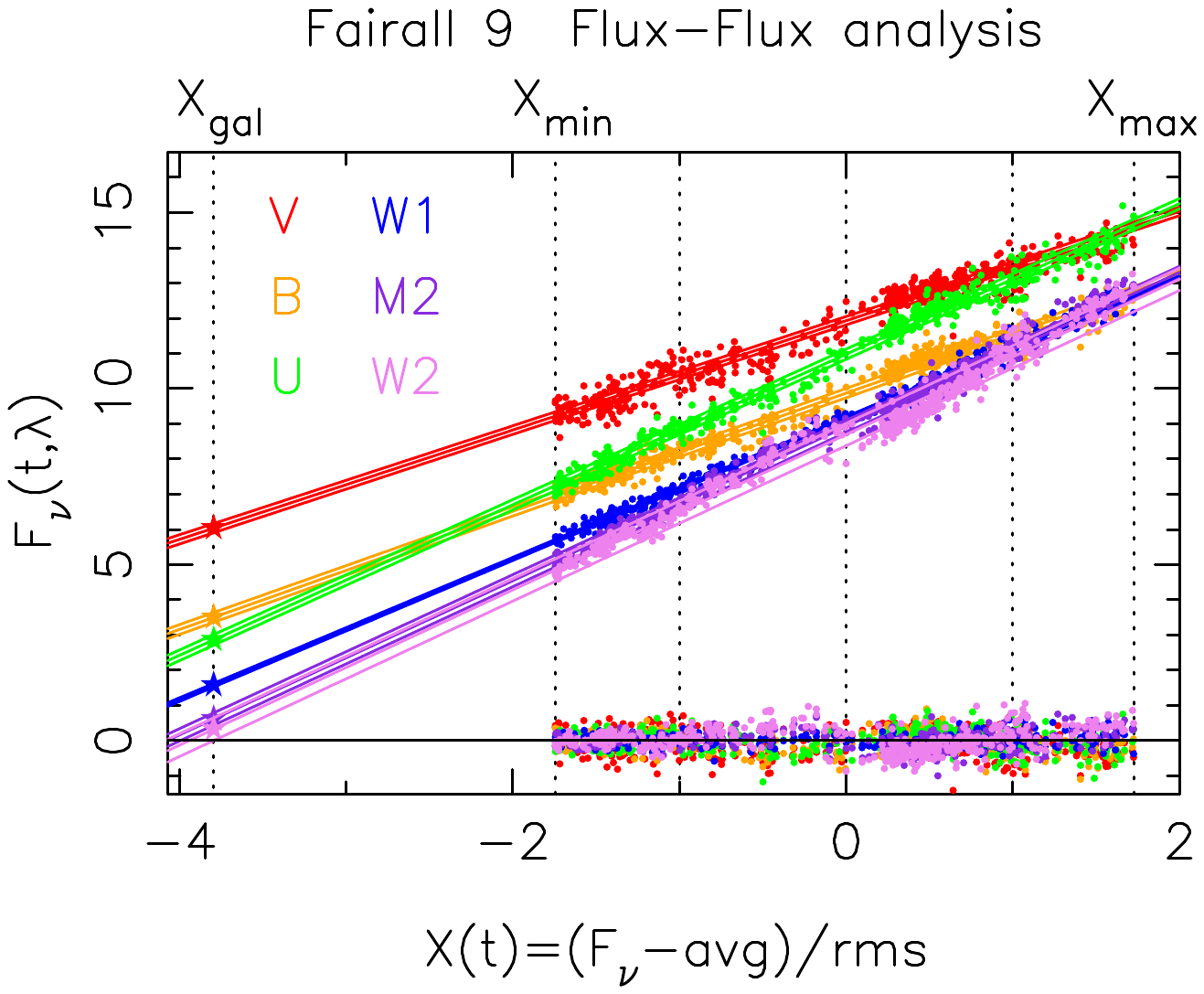} ~
  \includegraphics[trim=15mm 5mm 55mm 20mm,clip,width=0.49\textwidth]
 {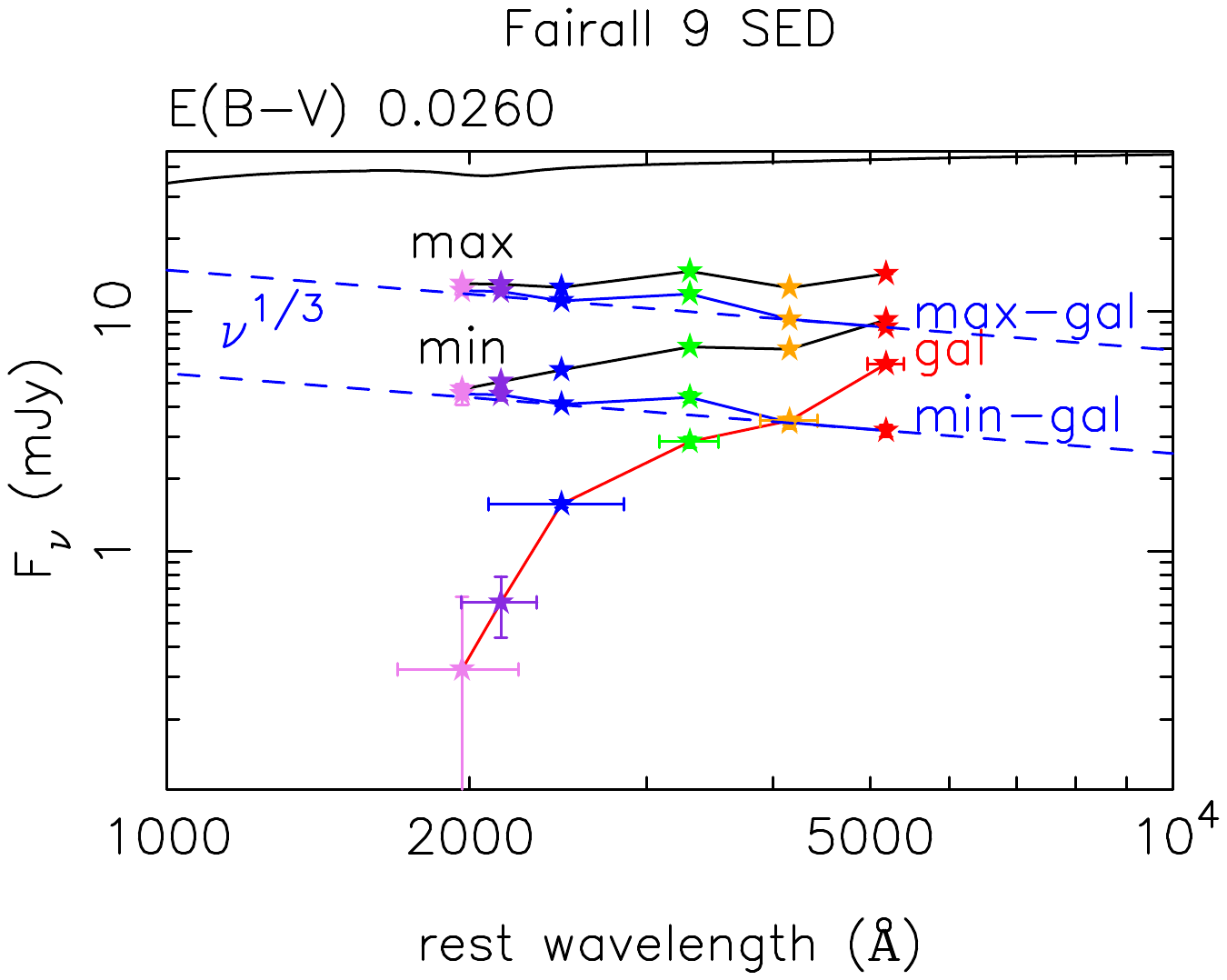} 
\caption{
Flux-flux analysis and resulting spectral energy distributions (SEDs) for the variable (disk) and constant (host galaxy) components of the AGN light.
{\it Left:} Observed fluxes in the 6 Swift bands are plotted against a dimensionless measure $X$ of the brightness, where $X=0$ corresponds to the mean and $X=\pm1$ the mean $\pm$ rms brightness. 
Best linear fits and $\pm 1\sigma$ uncertainties are shown for each band.
Extrapolating to fainter levels, define $X_{\rm gal}=-3.79$ where the UVW2 flux is 1$\sigma$ above zero. 
Evaluating the extrapolated fluxes at $X_{\rm gal}$ defines the constant galaxy SED $F_{\rm gal}(\lambda)$, and subtracting that leaves the variable disk flux $F_{\rm disk}(\lambda,X)$.
{\it Right:} SEDs corrected for Milky Way dust extinction (black), $F_{\rm gal}(\lambda)$ (red) for the host galaxy and $F_{\rm disk}(\lambda,X)$ (blue) for the variable disk at minimum and maximum brightness.
Power-law SEDs with $F_\nu\propto \nu^{1/3}$ (blue dashed) are shown for reference. 
Horizontal error bars on the galaxy SED indicate the \swift\ filter bandwidths.
}
\label{fig:ff}
\end{center}
\end{figure*}

Fig.~\ref{fig:ff} illustrates our flux-flux analysis of the six-band \swift\ data on F9. 
In the left panel, the observed fluxes in the light curves shown in Fig.~\ref{fig:lccf} are plotted against a dimensionless brightness level $X$, defined to be zero at minimum and unity at maximum brightness.
the dimensionless lightcurve shape $X(t)$, normalised to zero mean $\left<X\right>=0$ and unit variance $\left<X^2\right>=1$.
The flux-flux analyses assumes a linear model,
\begin{eqnarray}
    F(\lambda,X(t)) &=& F_0(\lambda) + F_1(\lambda) \, X(t)
\\    &=& F_{\rm gal}(\lambda) + F_{\rm disk}(\lambda)\, \left( X(t) - X_{\rm gal} \right)
\ ,
\end{eqnarray}
representing the total (disk+galaxy) light at different brightness levels.
The intercept $\bar{F}(\lambda)$ at $X=0$, and the slope $dF/dX=\Delta F(\lambda)$,  are the mean and rms of the variations at each wavelength.
the minimum flux $F_0(\lambda)$ at $X=0$,
and the increase in flux by $F_1(\lambda)$ at $X=1$.
Note that $X(t)$ is also modelled; it is not simply the observed X-ray light curve.)
The slopes in Fig.~\ref{fig:ff} are evidently steeper in the UV and become flatter at longer wavelengths, indicating that the variable component has a relatively blue SED, with $F_\nu$ decreasing with wavelength.
Bucking this trend, the $U$-band slope is steeper than the adjacent bands.
Extrapolating the linear model to fainter levels, $X<0$ $X<X_{\rm min}$, predicts the flux with the variable disk component at fainter levels than observed during the campaign. Because fluxes must be positive, we define 
 $X_{\rm gal}=-3.79$
 at the point where the $W2$ flux is $1\sigma$ above zero.
Evaluating the model fluxes at $X_{\rm gal}$ then gives an estimate of the constant (host galaxy) SED $F_{\rm gal}(\lambda)$, and subtracting this gives the variable (disk) SED $F_{\rm disk}(\lambda,X)$.

\begin{center}
\begin{deluxetable}{cccccccc}
    \tablecaption{Galaxy and Disk SEDs from the Flux-Flux Analysis in Fig.~\ref{fig:ff}. 
    \label{tab:ff}} 
\tablewidth{0.48\textwidth}
\tablecolumns{5}
\tablehead{
   \colhead{(1)} & \colhead{(2)} & \colhead{(3)} & \colhead{(4)} & \colhead{(5)} 
\\ \colhead{band}
    & \colhead{$\lambda_p$} 
    & \colhead{$F_{\rm gal}$}
    & \colhead{$F_{\rm disk, min}$}
    & \colhead{$F_{\rm disk, max}$}
\\  & \colhead{(\AA)}
    & \colhead{(mJy)}
    & \colhead{(mJy)}
    & \colhead{(mJy)} 
    }
\startdata
    $W2$ & 1965 
&  $0.32\pm0.32$  
& $4.52\pm0.45$ 
& $12.18\pm0.45$ 
\cr  $M1$ & 2147 
& $0.61\pm0.18$ 
& $4.49\pm0.24$ 
& $12.09\pm0.24$ 
\cr  $W1$ & 2467 
& $1.57\pm0.05$ 
& $4.10\pm0.06$ 
& $11.03\pm0.06$ 
\cr  $U$ & 3310 
& $2.86\pm0.15$ 
& $4.38\pm0.21$ 
& $11.80\pm0.21$ 
\cr  $B$ & 4158 
& $3.50\pm0.14$ 
& $3.44\pm0.19$ 
& $9.27\pm0.19$ 
\cr  $V$ & 5186 
& $6.04\pm0.13$ 
& $3.18\pm0.17$ 
& $8.57\pm0.17$ 
\enddata
 \tablecomments{
 Column~1: \swift\ band,
 Column~2: Rest-frame pivot wavelength,
 Columns~3-5: Flux densities, corrected for MW dust extinction E(B-V)=0.026~mag, for the constant host galaxy and for the variable accretion disk at the minimum and maximum brightness levels during the \swift\ campaign.
 }
\end{deluxetable}
\end{center}

The resulting galaxy and disk SEDs are detailed in Table~\ref{tab:ff} and
shown in the right-hand panel of Fig.~\ref{fig:ff}.
The observed flux densities are converted from $F_\lambda$ in Fig.~\ref{fig:lccf} to $F_\nu$ in Fig.~\ref{fig:ff} using pivot wavelengths computed from the sensitivity curves of the \swift\ passbands. 
The rest-frame pivot wavelengths are given in Table~\ref{tab:ff}.
This is advantageous because the disk SED is relatively flat in $F_\nu$.
The SEDs are also corrected for Milky Way dust extinction for 
$E(B-V) = 0.026\,{\rm mag}$ \citep{Schlafly2011}.

Note that the disk SED increases by a factor of 2.7 between the minimum and maximum levels observed during the campaign. 
It is remarkable that this relatively large increase occurs with little change in the SED of the variable component. 
The rms residuals to the fit in Fig.~\ref{fig:ff} are below 0.35~mJy in all six bands.
This linearity in the response across a wide range of UV and optical wavelengths is well documented in other Type~1 AGN, for example in NGC~5548 \citep{Starkey17} the disk SED remains unchanged over a factor 5 change in brightness.

While the observed disk+galaxy SED appears to be ``bluer when brighter,''
the linear model attributes this to a relatively red constant galaxy SED plus a relatively blue variable disk SED.
Alternatively, \cite{Korista01} suggest that as much as 1/3 of this effect may arise from the variable DCE contribution.
Note in Fig.~\ref{fig:ff} that the disk SED is close to the $F_\nu \propto \nu^{1/3}$ SED predicted for reprocessing on a steady-state thin blackbody accretion disk. 
Note also a significant excess flux in the $U$ band of about 16\% relative to the adjacent bands, indicating the size of the Balmer jump and thus the level of Balmer continuum emission in the correlated variations.

\section{Discussion}\label{sec:disco}

\subsection{Accretion disk + BLR model
}\label{sec:disk}

We use the available observations of F9 to construct a combined disk-BLR model for the source.
We assume that the UV/optical continuum SED 
is produced by an optically thick, geometrically thin accretion disk as calculated by \cite{Slone12}.
The disk properties and BH mass are adjusted to agree with earlier spectroscopic and X-ray measurements and with the observed flux in the $W2$ \swift\ band \citep{Hernandez20}.
This gives $M_{\rm BH}=2.6\times 10^{8} M_\odot$\; $L/L_{\rm Edd}=0.035$, $a=0.7$, and $\alpha_{ox}=1.4$, where {\it a} is the spin parameter.
Assuming face-on inclination results in $\lambda L_{\lambda} (5100\, {\text{\AA}}) = 10^{44}$~erg\,s$^{-1}$, but the luminosity seen by the BLR can differ by a factor of 1.5--2 due to the unknown disk inclination and the large amplitude continuum variations observed along our line of sight during the campaign.

\begin{figure}
\begin{center}
 \includegraphics[width=0.48\textwidth]{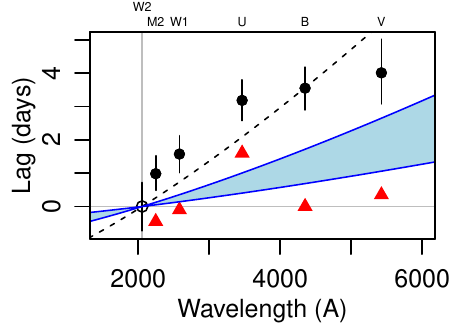} 
\caption{
Plot of lag vs. wavelength for the UVOT data in Table~\ref{tab:full}. 
The dashed line is a fit to $\tau \propto \lambda^{4/3}$ for the five longer wavelength UVOT CCF points, shown as filled circles.
(The $W2$ ACF, shown as an unfilled circle at 1928~\AA, does not participate in the fit because the fit is forced through $\tau=0$ for the ACF.)
These data are in the observed (not rest) frame with no correction for the small effects of time dilation.
In addition the blue shaded area is generated from the estimated black hole mass and Eddington ratio parameters, under the Wien assumption (upper bound) and flux-weighted assumption (lower bound).
The red triangles show interband lag predictions based on the \cite{Netzer22} BLR plus zero-lag disk assuming the same source parameters as in that paper  (H. Netzer, priv. comm.).
See text for further details.
\label{fig:taulambda}
}
\end{center}
\end{figure}

To further investigate the UV/optical variability, Fig.~\ref{fig:taulambda} shows the UV/optical lags $\tau$ and uncertainties as a function of the central wavelength $\lambda$ of each band.
The lag data measured relative to the $W2$ band are fitted with a power-law function of the form $\tau = \tau_0\, \left( (\lambda/\lambda_0)^{4/3} - 1 \right)$, where 
$\tau_0$ is the lag at the reference wavelength $\lambda_0=2050$~\AA\ (the center of the $W2$ band), and where $\tau_0$ is subtracted to force the lag to be zero at $\lambda_0$.  
The parameter $\tau_0$, interpreted as light travel time, measures the mean radius of the disk, weighted by its response at $\lambda_0$ to changes in the irradiation.
This fit, shown as a dashed line in Fig.~\ref{fig:taulambda}, formally yields a fitted value of $ \tau_0 = 1.81 \pm 0.20 $~lt-day, and although the lags generally do increase with wavelength, the overall fit is rather poor, with $\chi_\nu^2 = 7.9/4$ dof.

Next we use Eqn.~1 of \cite{Edelson19} to estimate the light-crossing radius $R$ of an annulus emitting at a characteristic wavelength $\lambda$, which yields:
\begin{equation}\label{eq:}
R = 0.09 \left(X\frac{\lambda}{1928{\text{\AA}} } \right)^{4/3}\nonumber M_8^{2/3} \left(\frac{\dot m_{\rm Edd}}{0.10} \right)^{1/3}
\textrm{lt-dy}
\end{equation}
where $X$ is a multiplicative scaling factor of order unity that accounts for systematic issues in converting the annulus temperature $T$ to wavelength $\lambda$ at a characteristic radius $R$, $M_8$ is the black hole mass in units of $10^8 \,{\rm M}_\odot$ and $\dot{m}_\mathrm{Edd}$ is the Eddington ratio $ L_\mathrm{bol} / L_\mathrm{Edd} $. 
Note that this assumes a face-on disk.
If we assume an annulus of radius $R$ emits at an observed wavelength corresponding to the temperature given by Wien's Law, then $ X = 4.97 $.
If instead the flux-weighted radius is used, then $ X = 2.49 $.
(The flux-weighted estimate assumes that the temperature profile of the disk is described by $ T \propto R^{-3/4}$; \citealt{Shakura73}.)
In both the Wien and flux-weighted cases, the disk is assumed to have a fixed aspect ratio $H/R$ and to be heated internally by viscous dissipation and externally by the coronal X-ray source extending above the disk.
These two assumptions are used to generate the blue shaded area in Fig.~\ref{fig:taulambda}, where the Wien assumption gives the upper bound and the flux-weighted assumption yields the lower bound.
The estimated disk size overshoots that derived under the Wien assumption by a factor of 1.8 and the flux-weighted assumption by 4.5.

In summary, the lag spectrum points to an accretion disk that is somewhat larger than expected, even discounting the $U$ band, where DCE from the BLR clearly contributes to the large net lag. 
We note again, however, that the systematic uncertainties are quite large. 
We therefore consider below the possibility that {\em all} of the interband lags are attributable to DCE. 
Under this hypothesis, the interband lags from the 
accretion disk are too short for us to measure and the continuum light curves for the various bands are, for all practical purposes, assumed to be identical. 
In the next section, we model DCE from the BLR to test this hypothesis.

\subsection{BLR diffuse continuum and emission line emission}\label{sec:blr}

Recently \cite{Netzer22} analyzed IBRM data from F9 as well as other AGN and finds that the entire lag spectrum can be explained as emission from the BLR (DCE plus line emission) and a ``zero-lag'' disk (one that is so small that the interband lags are not measurable with current data). 
In Fig.~\ref{fig:taulambda} the red triangles show the result of using the \cite{Netzer22} model to estimate the emission-weighted mean distance from the central source to the emitting region, as a function of wavelength and then convolved with the \swift\ bands, for comparison with the observed lags (black dots with error bars).
Note that the $M2$ lag is predicted to be strongly negative relative to $W2$ in this model.
The $M2$ lag is smaller than the $W2$ lag in all realizations using the original input source parameters of the \cite{Netzer22} paper.
Fig.~\ref{fig:bands} shows why: a complex of strong BLR emission lines (e.g., C{\sc IV} 1550\AA, C{\sc III}] 1909\AA) falls in the observed $W2$ band, but the $M2$ band contains much less line emission.
This substantial BLR emission increases the mean emissivity distance and thus the expected W2/M2 lag, in the same fashion as Balmer jump (continuum) emission from the BLR produces excess U band lags.
Thus the model lags due to BLR line plus DCE emission alone are always larger for $W2$ than for $M2$.
The addition of a zero-lag (or near zero-lag) disk, no matter how strong its variability amplitude is compared to that of emission from the BLR, may affect the magnitude of this interband lag relationship, but cannot change its direction.

We also note that similar behavior has been observed in two earlier IBRM campaigns: early \iue\ monitoring of NGC~7469 finds that 1350~\AA\ continuum leads 1825~\AA\ by $\sim$0.25 day \citep{Wanders97} and the 2014 \swift/\hst\ campaign on NGC~5548 shows small but significantly increasing lags throughout the UV \citep{Fausnaugh16}.
Most other IBRM campaigns have lacked the sensitivity and/or temporal resolution to characterize significant lags within the UV, but it is worth noting that of the three for which such lags have been detected, all have been positive (in the sense that the shorter wavelengths lead the longer wavelengths).

Thus we conclude that the observed \swift\ lag spectrum of F9 and likely the other two AGN cannot be completely explained by the BLR plus zero-lag disk model of \cite{Netzer22}.
We note however that for this particular model of the BLR there is a great deal of freedom, so for instance if the Fe\,{\sc ii} emission (a notoriously difficult feature to characterize) is much stronger than used in the original model, the $M2$ and $W1$ lags would become larger relative to $W2$, and could even be positive as seen in the data.
Further modeling is planned to investigate this and other possibilities.

\begin{figure*} \centering    
\includegraphics[width=1.1\textwidth]{
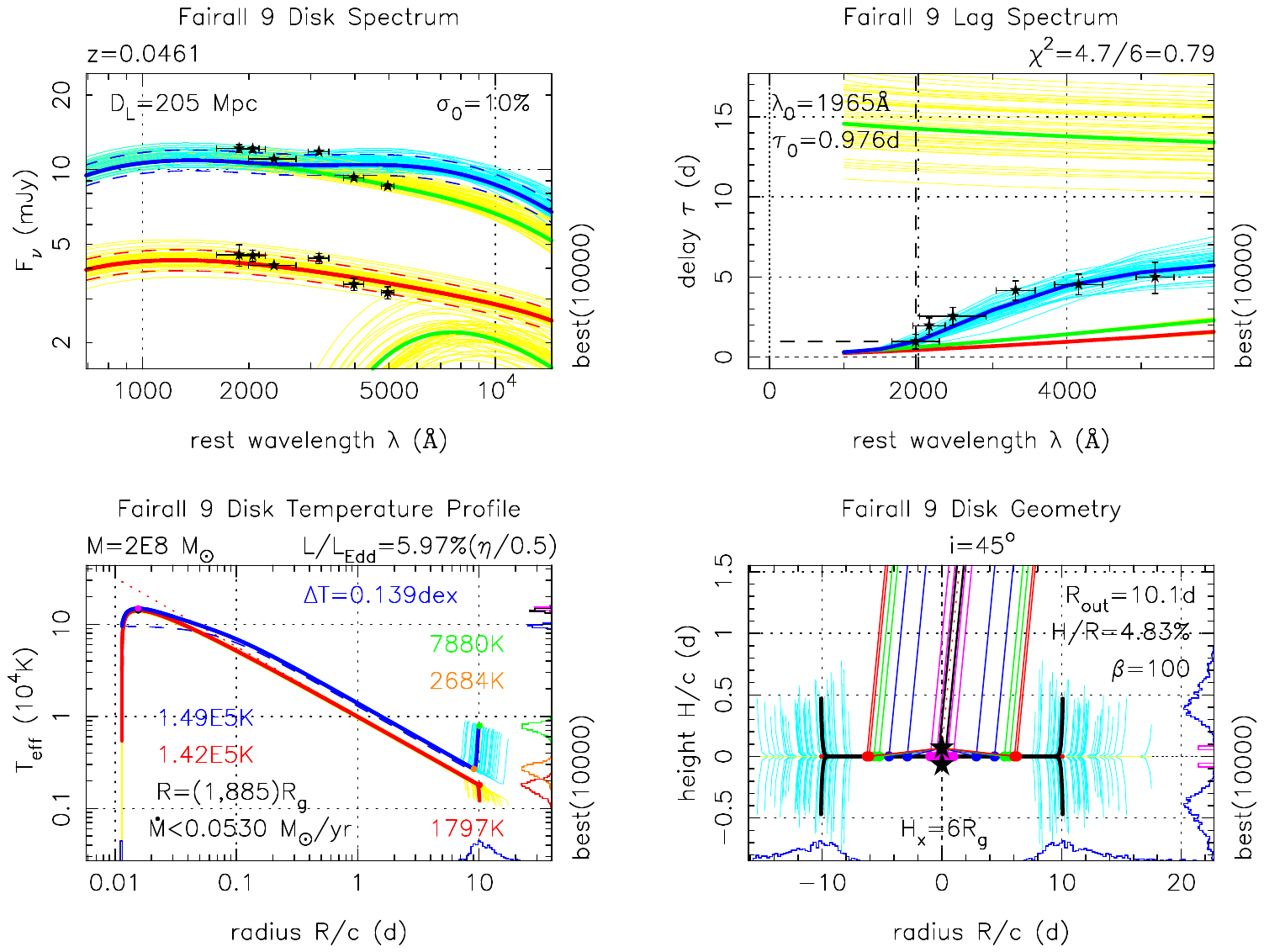
}
\caption{Bowl model predictions fitted to the six-band CCF lags (top right) and two six-band AGN SEDs (top left) at the minimum and maximum levels observed during the campaign. Blue and red curves give the model predictions with the lamp-post irradiation on and off, respectively. 
The model thickness profile $H(R)$ (lower right) has a high power-law index, giving a flat disk with a steep rim near $R_{\rm out}/c\sim10$~days.
Here colored lines depict light rays from the lamp-post down to the disk and up to the observer at $i=45^\circ$. Note the aspect ratio is $H/R\sim5$\% at the crest of the rim.
The temperature profile $T(R)$  (lower left) balances blackbody cooling and the sum of internal viscous and external irradiative heating. The flat disk has $T\propto R^{-3/4}$, falling from $10^5$~K to $2000$~K and then rising to $8000$~K at the crest of the steep rim.
\kdh{
Green curves in the upper panels show SEDs and lags for the flat disk and steep rim components separately.
Yellow and cyan curves for 30 random MCMC samples indicate the fit uncertainty.
Histograms on the lower and right axes of the lower panels give posterior distributions for 5 fiducial temperatures, inner and outer radii, and thickness at the rim.}
\label{fig:bowl}}
\end{figure*}

\subsection{The accretion disk ``bowl'' picture}\label{sec:bowl}

\rae{
The lag spectrum in Fig.~\ref{fig:taulambda} clearly rises in the UV and then appears to flatten at longer wavelengths, albeit with some uncertainty.
This convex lag spectrum shape is commonly seen in the IBRM lag spectra of other AGN as well \cite[e.g.][]{Edelson15, Hernandez20}.
In a picture built around a standard \citep{Shakura73} flat accretion disk, one obvious way to produce such a lag spectrum is to introduce an additional component of ``piled-up'' material at an outer edge of the disk.
}

\rae{
Thus in this section we fit these data with a ``bowl’’ model \citep{Starkey23} that assumes lamp-post irradiation and 100\% blackbody reprocessing on the disk surface, but relaxes the assumption that the disk thickness profile $H(R)$ is negligible and constant for all radii.
This allows us to introduce a ``wall’’ of height $H_{\rm out}$ at radius $R_{\rm out}$, which is associated with the opening angle $H_{\rm out} / R_{\rm out}$.
This wall could represent material in the BLR, a wind, a dusty torus, or (most likely) a combination of these three physical components.
Additionally we assume a lamp-post irradiator at a height $H_{\rm x}$ above the disk and that the disk itself is truncated at the inner radius $R_{\rm in}$.
We also assume a fixed inclination angle $i=45^\circ$, luminosity distance $D_{\rm L}=205$~Mpc, from the redshift $z = 0.046145$ and $\Lambda$CDM cosmology $(h_0,\Omega_\Lambda, \Omega_M)=(0.7,0.7,0.3)$, and black hole mass $M=2.0\times10^8$~M$_\odot$.
}

\rae{
This yields an initial model with seven parameters: 
$\dot{M}$, 
$R_{\rm in}$, 
$R_{\rm out}$, 
$H/R$, 
$\Delta L_{\rm x}$, 
$H_{\rm x}$, and
the model SED uncertainty
$\sigma_0$.
Note that because the current dataset has approximately daily sampling, 
it has very little power to constrain the parameters $R_{\rm in}$ and $H_{\rm x}$, as these distances correspond to light-travel times of less than an hour.
Thus we fix these at $R_{\rm in}=R_{\rm g}$ $H_x=6\,R_{\rm g}$, corresponding to a high prograde black hole spin and a moderate X-ray corona height. 
Our fits prefer such small values, but we emphasise that these are unreliable parameters for this analysis.
(These inner-disk parameters are best probed with {\it XMM} long-looks, e.g. \cite{Fabian13}.)
}

\rae{
Fig.~\ref{fig:bowl} presents a bowl model fit of the five remaining free parameters ($\dot{M}$, $R_{\rm out}$, $H/R$, $\Delta L_{\rm x}$, and $\sigma_0$) to the six lags measured in the cross-correlation analysis (Table~\ref{tab:full}), and simultaneously to the minimum and maximum AGN SEDs from the flux-flux analysis (Table~\ref{tab:ff}), a total of $3\times6=18$ data constraints.
The bowl model shows rms residuals of $\sigma_0\sim10$\,\% when fitted to the observed bright and faint SED.
An upper limit to the accretion rate, $\dot{M}<0.053\,M_\odot$\,yr$^{-1}$, is set by the faint disk SED, since irradiation also contributes to heating, and viscous timescales are too long to account for the observed variations.
With the disk SED being $F_\nu\propto\nu^{1/3}\propto T^{1/3}$,
the observed bright/faint flux ratio $\approx2.7$ requires a rise in $\log{T}$ of $\log{(2.7)}/3=0.14$\,dex due to increased irradiation.
The wall at $R_{\rm out}/c \sim 10$\,d, has a lag $(R_{\rm out}/c)(1+ (2/3)\cos{i})\approx 14$\,d.
The wall is implemented as $H\propto R^{100}$.
Irradiation heats the wall from $\sim2000$\,K at its base to $\sim8000$\,K at its crest.
The resulting wavelength-dependent mix of short disk lags and longer wall lags matches the size and convex shape of the observed lag spectrum.
The wall is not tall, $H/R\sim5$\%,
and it sits near to where the
disk temperature, falling as $T\propto r^{-3/4}$ , crosses dust sublimation temperatures. 
The disk likely extends to larger radii where dust grains can survive in the cooler disk atmosphere. Here we can expect an increase in thickness \citep{Baskin18} and launching of a failed radiatively accelerated dusty outflow \citep[FRADO;][]{Czerny11, Naddaf21}.
The bowl model's wall may thus correspond to the transition from a dusty disk atmosphere launching BLR clouds, to a dust-free disk atmosphere inside the inner edge of the BLR.
A similar picture, with dust-free gas being launched from the outer disk to form the inner edge of the BLR may also be seen in NGC~4395 \citep{McHardy23}.
}

\subsection{X-ray/UV relationship}\label{sec:xrays}

Probably the most surprising result of the early \swift\ IBRM experiments was that, while UV and optical variations were always highly correlated with lag spectra that followed the predicted $ \tau \propto \lambda^{4/3} $ relation, the X-ray to UV correlations were much weaker and the interband relationship was unclear.
In some cases the X-rays and UV appear strongly related, especially after detrending of the UV \cite[e.g.,][]{Hernandez20,Lawther23} while in others the relationship is weaker or non-existent \cite[e.g.,][]{Edelson17}.

The sheer size of this F9 dataset allows deeper investigation of this complex relationship.
The quarterly analysis shown in Fig.~\ref{fig:quarterly} indicates that the lags within the UV/optical are highly stable, but the X-ray to UV lags vary in magnitude and in some cases even sign from one quarter to the next, in the same object.
This strengthens the view that the observed 0.3-10 keV X-ray variability is not the sole driver of the observed UV/optical variability.
Instead, the driver must either be obscured from our line of sight, or it must be in a band that is either higher or lower in energy than observed by \swift, or (the most extreme possibility) the entire reprocessing model could be invalid.

\section{Conclusions}\label{sec:concl}

This F9 campaign is the most extensive multiband AGN monitoring campaign undertaken up to this point in time (MJD 58900), although an ongoing campaign on Mrk 817 should eventually exceed this \cite{Cackett23}.
We perform a systematic analysis of the impact of detrending on the standard cross-correlation analysis and find that in this case a second order polynomial function provides the most accurate measurement of interband lags.
This methodology is easily applicable to other IBRM datasets, and because it 
offers and objective criterion for selecting the detrending scheme without user intervention, it could lessen the likelihood of controversial results, e.g. the widely disparate X-ray/optical lags claimed in the same {\em IUE/ASCA} observations of NGC~7469 (\citealt{Nandra00}, \citealt{Pahari20}).
Future papers on these data will systematically investigate detrending in even greater detail, for instance studying Gaussian and boxcar smoothing and comparing the results with {\tt JAVELIN} (Peterson et al., in prep.).

This large dataset provided the clearest detection to date of positive interband lags within the UV, with significant detections between the \swift\ $W2$ and both $M2$ and $W1$ bands.
The presence of strong BLR line emission in $W2$ means that this lag spectrum cannot be explained by a combination of emission from the BLR and a small disk.
Likewise the excess $U$-band lag seen in this campaign as well as almost all previous IBRM campaigns indicates that the full lag spectrum cannot be the result of a pure thin disk of any size.
Thus we conclude that neither BLR DCE, a thin accretion disk, nor a combination of both can fully account for the observed lag spectrum.
At this time the most promising theoretical approaches to solving this problem involve truncated disks \cite[e.g.,][]{Hagen23,McHardy23} or reprocessing without reference to an X-ray corona \cite[e.g.,][]{Neustadt22}.

Along these lines, Fig.~\ref{fig:bowl} shows the result of fitting the F9 lags and UV/optical SEDs with a `bowl' model that invokes lamp-post irradiation and blackbody reprocessing on a power-law disk thickness profile $H\propto R^\beta$, as introduced in the context of fitting the UV/optical variations of NGC~5548  \citep{Starkey23}. 
We find that \del{blackbody reprocessing on a flat disk with a steep outer rim does reproduce the observed lags that rise steeply in the UV and flatten off in the optical, while roughly matching also the SED of the UV/optical variations. 
Moreover,} the observed lags imply that this model's steep rim occurs \rae{at a distance ($\sim$10 lt-dy)} where the flat disk temperature profile crosses the dust sublimation temperature, in agreement with the predictions of models that invoke dust opacity to increase the disk thickness \citep{Baskin18} and/or to launch a failed radiatively accelerated dusty outflow \citep{Czerny11,Naddaf21} that has been proposed as the origin of the low-ionization component of the BLR.

Finally we note that extensive ground-based photometry and spectroscopy and high throughput NICER X-ray monitoring was also gathered for F9 throughout this period.
The analysis of those datasets should further elucidate the role of the disk and BLR in producing the observed relations between variations in different bands. 
Those data and results will be reported in future papers.

\begin{acknowledgments}
The authors thank the Neil Gehrels \swift\ Observatory team for their tireless and successful efforts to execute this program.
We thank the anonymous referee for constructive criticism of an earlier version of this paper.
RE and JMG gratefully acknowledge support from NASA under \rae{the ADAP award 80NSSC17K0126}.
JMG also gratefully acknowledges support from NASA under the Swift awards 80NSSC22K1492 and 80NSSC24K0513.
IMcH thanks the UK Science and Technology Facilities Council for support under grant ST/J001600/1.
MV gratefully acknowledges financial support from the Independent Research Fund Denmark via grant number DFF 8021-00130.
\end{acknowledgments}

\software{{\tt HEAsoft} (v6.22.1; Arnaud 1996), {\tt FTOOLS} (Blackburn 1995), {\tt sour} (Edelson et al. 2017)}

\end{document}